\documentstyle[multicol,overcite,pre,aps,epsf]{revtex}
\newcommand{\wdtxt}{\end{multicols}\widetext\hspace{-\parindent}\hrulefill 
\hspace{2.0in}\mbox{}}
\newcommand{\nrtxt}{\mbox{}\hspace{2.0in}\hrulefill\mbox{}\begin{multicols}{2}
\narrowtext\hspace{-\parindent}}
\begin{document}		    
\draft

\title{Perturbation Theory for the Breakdown of Mean-Field Kinetics 
in Oscillatory Reaction-Diffusion Systems}

\author{Mikhail V. Velikanov and Raymond Kapral\\ 
Chemical Physics Theory Group, Department of Chemistry,\\
University of Toronto, Toronto, Ontario M5S 3H6, Canada.}

\maketitle

\begin{abstract}
	Spatially-distributed, nonequilibrium chemical systems described by a 
Markov chain model are considered. The evolution of such systems arises 
from a combination of local birth-death reactive events and random walks 
executed by the particles on a lattice. The parameter $\gamma$, the ratio of 
characteristic time scales of reaction and diffusion, is used to gauge 
the relative contributions of these two processes to the overall dynamics. 
For the case of relatively fast diffusion, i.e. $\gamma \ll 1$, an approximate 
solution to the Markov chain in the form of a perturbation expansion in powers 
of $\gamma$ is derived. Kinetic equations for the average concentrations that 
follow from the solution differ from the mass-action law and contain memory 
terms. For a reaction-diffusion system with Willamowski-R\"ossler reaction 
mechanism, we further derive the following two results: a) in the limit of $\gamma 
\rightarrow 0$, these memory terms vanish and the mass-action law is recovered; 
b) the memory kernel is found to assume a simple exponential form. A comparison 
with numerical results from lattice gas automaton simulations is also carried 
out. 
\end{abstract}


\begin{multicols}{2}
\narrowtext

\section{Introduction}
	The dynamics of chemical reactions in condensed media is commonly 
described by a set of reaction-diffusion equations \cite{gen} of the following 
general form:

\begin{equation}
{\partial \overline{{\bf x}}({\bf r},t) \over \partial t} = 
{\bf R}(\overline{{\bf x}}({\bf r},t)) + {\bf D} \nabla^2 \overline{{\bf x}}({\bf r},t),
\label{RD}
\end{equation}
where $\overline{{\bf x}}({\bf r},t)$ is vector of concentrations, ${\bf r}$  
position vector, $t$ the time; ${\bf R}(\overline{{\bf x}}({\bf r},t))$ denotes 
mass-action law terms, and ${\bf D}$ is the matrix of diffusion coefficients. 			      

	Such a description, while able to capture some of the features of 
the system's evolution, is not entirely free of shortcomings. For instance, 
as can be easily seen from (\ref{RD}), rescaling the diffusion coefficients
by a constant factor is equivalent to rescaling of length. Contrary to this 
observation, the reaction dynamics in spatially-distributed systems varies 
with diffusion in a much more complex way; thus, it has been shown that in 
the limit of slow diffusion reaction dynamics has pronounced memory character 
\cite{sokolov,felderhof}, while in the case of very fast diffusion it closely 
follows the mass-action law. Significant progress has been made in understanding
the dynamics of simple diffusion-limited chemical reactions (e.g. $A + B 
\rightarrow \emptyset$) in low-dimensional systems~\cite{privman}. 
Also, Chapman-Enskog methods have been applied to multi-variate master 
equations to study reactive correlations in simple reaction-diffusion 
systems. \cite{broeck} However, to
date the dynamics of more complex reaction-diffusion systems whose mass-action 
laws exhibit periodic and chaotic oscillations has received much less attention.
Recent numerical studies on such systems using lattice gas automaton models~\cite{wu} 
and master equations~\cite{baras,nicolis} have indicated that there are interesting 
effects pertaining to modification of the structure of oscillatory and chaotic 
attractors with decreasing rate of diffusion. 

	The purpose of this paper is to develop a statistical mechanical 
theory that approximately accounts for the influence of finite diffusion 
and reactive correlations on the dynamics of nonequilibrium oscillatory 
chemical systems. We use a discrete-space, discrete-time probabilistic model 
defined on a set of sites, or nodes of a lattice. Its evolution is due to a 
combination of local reactive events and random walks which are executed by 
the particles of reactants on the lattice.  

	Our approach is based on a Chapman-Enskog-like development~\cite{uhlenbeck} 
applied to a Markov chain model. The expansion parameter $\gamma$ is the ratio 
of characteristic time scales of diffusion and reaction processes, small for
the diffusion-dominated regimes. We show that the corresponding kinetic equations 
for spatially-averaged concentrations involve memory terms. As an example, we 
apply this theory to Willamowski-R\"ossler reaction-diffusion system \cite{wr}. 
We find that the memory kernel for this system assumes simple exponential form. 
The memory kernel should have the same functional form for any reaction-diffusion
system whose reaction mechanism involves mono- and bimolecular steps only.
The dynamics predicted by the kinetic equations indicates that the influence of diffusion 
essentially amounts to a {\it backward shift} in the bifurcation cascade of 
the mass-action law. We compare these theoretical predictions with lattice gas 
automaton simulations of Willamowski-R\"ossler model, and find that they are 
qualitatively in agreement. 

	The paper is organized as follows. In Sec. \ref{model} we introduce 
the Markov chain model and construct its evolution operator. In Sec. \ref{theory} 
we present a perturbation expansion for the probability distribution function
around the local binomial form and obtain equations that determine evolution 
of the terms of this expansion. The perturbed distribution function incorporates
the local correlations between different species induced by the reactive events
and thus cannot be factorized into a product of single-species distributions.
The derivation of kinetic equations for global average concentrations (i.e. local 
concentrations averaged over the volume of the system) concludes this section. 
In Sec. \ref{WR} we derive the memory kernel for Willamowski-R\"ossler 
reaction-diffusion system and compare the predictions of the theory with the 
results of lattice gas automaton simulations. Finally, Sec. \ref{conclusions} 
contains discussion of our results.
 
\section{Markov Chain Dynamics}
\label{model}

	The system under consideration consists of a number of chemical species 
diffusing and reacting in solution. We assume a discrete-space, discrete-time 
description of the system by partitioning the space into cells of volume $V$ 
and time into intervals of arbitrary length. In the following, we view each 
cell as a node on a lattice and operate within this lattice representation. The solvent 
comprises a chemically inert species so that its only effect is to maintain the 
random walk dynamics of the solute particles. Furthermore, we associate $N$ 
distinct ``channels'' with every cell; each channel can be occupied by no more 
then one particle of each solute species. Thus, if the number of solute species 
is $\nu$, the number of solute particles in each cell cannot exceed $\nu N$ -- 
this is the so-called exclusion principle. At any given moment of time, the 
state of the system is fully specified by a set of state variables -- the 
number and chemical nature of the particles occupying every cell. The system 
evolves by transition to one of the accessible states once every time 
interval with probability determined by the initial state. From the statistical 
point of view the system is described by the probability distribution function 
$P({\bf s},t)$ defined on the space of discrete states of the system, whose 
time dependence is due to a Markov chain \cite{cox}, i.e.:
\wdtxt
\begin{equation}
\label{chain}
P({\bf s},n+1) - P({\bf s},n)  = \sum_{{\bf s}'}^{}
[T_{{\bf s}{\bf s}'} P({\bf s}',n) - T_{{\bf s}'{\bf s}} P({\bf s},n)],
\end{equation}						      
\nrtxt
where ${\bf s}$ is a set of state variables, $n$ is the integer time variable, 
and the summation extends over all possible states. Clearly, due to discreteness 
of the state variables, eq. (\ref{chain}) can be also written in matrix form:
\begin{equation}
\label{mchain}
P_{{\bf s}}(n+1) - P_{{\bf s}}(n) = \sum_{{\bf s}'}^{} W_{{\bf s}{\bf s}'} 
P_{{\bf s}'}(n), 
\nonumber 
\end{equation}
where $W_{{\bf s}{\bf s}'}$ are the matrix elements of evolution operator, 
defined in terms of transition probabilities per unit time as follows:
\begin{equation}
\label{E-op}
W_{{\bf s}{\bf s}'} = T_{{\bf s}{\bf s}'} - \delta_{{\bf s}'{\bf s}} 
\sum_{{\bf s}''}^{} T_{{\bf s}''{\bf s}}.
\end{equation}	    
Here $\delta_{{\bf s}'{\bf s}}$ is a Kronecker delta.
						 
Henceforth, all vector quantities, such as state variables or concentrations, 
are denoted by bold lowercase letters, while bold uppercase letters stand for 
matrix quantities. Components of vectors and matrices are specified where 
necessary by lower indices. We designate operators by a hat over corresponding 
symbols, and average quantities by an overbar. Angle brackets will often be 
used as a shorthand notation for summing over the ensemble of state variables.

	For reaction-diffusion systems, transitions between different states 
occur due to two competing processes: passive transport of particles between 
cells and reactive collisions which change the nature of the particles within 
cells according to the reaction mechanism. Correspondingly, the evolution 
operator for the present model can be written as a sum of two terms, one for
the diffusion process and the other for reactions. Below we derive the particular
form of reaction and diffusion evolution operators to be used in our study.

\subsection{Diffusion Markov chain}
	
	Let the particles at every node be randomly distributed among $N$ 
channels. We pose a diffusion rule according to which the particles that have 
identical chemical nature and belong to the same channel propagate synchronously 
and in the same direction between neighboring nodes. The choice of a particular 
channel, direction of propagation, and species to be propagated is made at 
random each time the diffusion evolution operator is applied. We also impose 
periodic boundary conditions which, together with the above, completely 
characterizes transport of matter within the system. 

	A similar diffusion model was recently used in simulations of FitzHugh-Nagumo 
system \cite{malevan}. It was shown that the occupancies of individual nodes become statistically independent of each other 
after a short period of relaxation. Hence, for long-time behaviour, correlations 
between nodes are negligible, and the full probability distribution can be factorized
into a product of single-site distributions. This reduction of description 
is readily extended to the present model; specifically, one has for non-vanishing 
reduced transition probabilities:
\begin{eqnarray}	    
\label{no_corr}
T^D_{x_i({\bf r}) - 1,x_i({\bf r})} &=& {x_i({\bf r}) \over \nu N} \left(1 - 
{\chi_i({\bf r},n) \over N}\right), \nonumber \\
T^D_{x_i({\bf r}) + 1,x_i({\bf r})} &=& {\chi_i({\bf r},n) \over \nu N} 
\left( 1 - {x_i({\bf r}) \over N}\right), 
\end{eqnarray}
where
\begin{eqnarray}
\chi_i({\bf r},n) &=& {1 \over m} \sum_{{\bf r}' \in \cal N({\bf r})}^{} \;
\sum_{{\bf x}({\bf r}')}^{} x_i({\bf r}') P({\bf x}({\bf r}'),n) \nonumber \\ 
&=& {1 \over m} \sum_{{\bf r}' \in \cal N({\bf r})}^{} \overline{x}_i({\bf r}',n), 
\nonumber
\end{eqnarray}
$m$ is the coordination number of the lattice and $\cal N({\bf r})$ is 
neighbourhood of ${\bf r}$. Thus, $\chi_i({\bf r},n)$ is an average of the
mean occupation number of the $i$-th species over the immediate neighbourhood of
node ${\bf r}$. The factors in the first expression in (\ref{no_corr}) simply 
reflect the fact that the probability of a diffusive hop of a particle of the 
$i$-th species from the node ${\bf r}$ to a neighbouring node is given by a 
product of the probability that the $i$-th species is chosen for the hop,
$1/\nu$, the probability that a chosen channel is occupied at the node ${\bf r}$,
$x_i({\bf r})/N$, and the probability that the same channel is empty at the 
neighbouring node, $1 - \chi_i({\bf r},n)/N$. A similar interpretation can be 
also given to the second expression in (\ref{no_corr}).

	The reduced probability distribution $P({\bf x}({\bf r}),t)$ evolves 
according to the following equation:
\begin{eqnarray}
\label{Diff_chain}
P({\bf x}({\bf r}),n&+&1) - P({\bf x}({\bf r}),n) = \hat{W}^D P({\bf x}({\bf r}),n) 
\nonumber \\
&=& \sum_i^{} \hat{W}^{D^i} \left(\prod_{j \neq i}^{} 
\delta_{x'_j({\bf r}),x_j({\bf r})} \right) P({\bf x}({\bf r}),n),
\end{eqnarray}
with matrix elements of $\hat{W}^{D^i}$ given by (cf. (\ref{no_corr}) 
and (\ref{E-op})):
\wdtxt
\begin{eqnarray}
\label{Wdiff}
W^{D^i}_{x_i({\bf r}),x'_i({\bf r})} &=& {\chi_i({\bf r},n) \over \nu N}
\left(1 - {x'_i({\bf r}) \over N} \right) \delta_{x'_i({\bf r}),x_i({\bf r}) - 1}
+ {x'_i({\bf r}) \over \nu N} \left(1 - {\chi_i({\bf r},n) \over N}\right)
\delta_{x'_i({\bf r}),x_i({\bf r}) + 1} \\
&-& \left[{\chi_i({\bf r},n) \over \nu N} \left(1 - {x'_i({\bf r}) \over N} 
\right) + {x'_i({\bf r}) \over \nu N} \left(1 - {\chi_i({\bf r},n) \over N} 
\right) \right] \delta_{x'_i({\bf r}),x_i({\bf r})}. \nonumber
\end{eqnarray}
\nrtxt						       
Thus, the structure of diffusion Markov chain (\ref{Diff_chain}) arises 
naturally from the diffusion rule posed in the beginning, the
existence of a
maximum occupation number at any node (i.e. the exclusion principle), and the
reduction of description to the level of single-node distribution functions.

	Kinetic equations for average occupancies (concentrations) can be 
obtained in the standard way, by multiplying both sides of (\ref{Diff_chain}) 
by $x_i({\bf r})$ and summing over all possible states. One finds, after a few
simple transformations:
\wdtxt
\begin{eqnarray}		
\label{discr_diff}
\overline{x}_i({\bf r},n+1) - \overline{x}_i({\bf r},n) 
&=& {1 \over \nu m N} \sum_{{\bf r}' \in \cal N({\bf r})}^{} \left[
\overline{x}_i({\bf r}',n) - \overline{x}_i({\bf r},n) \right] \nonumber \\
&=& D \: \Delta \overline{x}_i({\bf r},n),
\end{eqnarray}
\nrtxt
where $\Delta$ is the discrete Laplacian operator. This is just a discretized 
diffusion equation with diffusion coefficient $D = (\nu m N)^{-1}$ (the same 
for all species). 

\subsection{Reaction Markov chain}
	The probabilities of reactive transitions are defined similarly to those 
in lattice gas automaton models. The details of the derivation for the models 
with exclusion principle can be found in~\cite{LGA}; here we only briefly 
summarize the general ideas and give the final results. 

	We consider reactions which are strictly local, i.e. they involve only
particles occupying the same node. Let the overall reaction mechanism consist
of $r$ elementary steps of the form:
\wdtxt
$$
n^f_{i,1} X_1 + n^f_{i,2} X_2 + \dots + n^f_{i,j} X_j + \dots 
\mathrel{\mathop{\kern0pt {\rightleftharpoons}}\limits^{{k_i}}_{k_{-i}}}
n^r_{i,1} X_1 + n^r_{i,2} X_2 + \dots + n^r_{i,j} X_j + \dots, \nonumber
$$    
\nrtxt
where the indices $i = 1, \dots, r$ and $j = 1, \dots, \nu$ label steps and
species, respectively. The transitions due to the $i$-th step in the mechanism are
characterized by two sets of stoichiometric coefficients $\{n^f_{i,1}, 
\dots, n^f_{i,\nu}\}$ and $\{n^r_{i,1}, \dots, n^r_{i,\nu}\}$. Specifically, if 
we define the vector ${\bf m}^{(i)}$ with elements \mbox{$m^{(i)}_j = n^r_{i,j} - n^f_{i,j}$}, 
the forward reaction at node ${\bf r}$ can be represented in terms of 
occupancies by:

$$ {\bf x}({\bf r}) \rightarrow {\bf x}({\bf r}) + {\bf m}^{(i)}. $$
Similarly, for the reverse reaction:

$$ {\bf x}({\bf r}) \rightarrow {\bf x}({\bf r}) - {\bf m}^{(i)}. $$
The transition probabilities of these birth-death reactive events can be derived 
using a simple combinatorial argument. That is, one expects that the probability 
of reaction must be proportional to the number of ways in which the particles 
of the reactants can be combined in order for that reaction to occur, given 
the present state of the node. Specifically, we put:
\wdtxt
\begin{eqnarray}
\label{T_reac}
T^R_{{\bf x}({\bf r})+{\bf m}^{(i)},{\bf x}({\bf r})} &=& h k_i \prod_{j=1}^{\nu} 
N^{n^f_{i,j}} {\left(N - n^f_{i,j}\right)! \over N!}\ {x_j({\bf r})! \over 
\left(x_j({\bf r}) - n^f_{i,j}\right)!}\ , \nonumber \\
T^R_{{\bf x}({\bf r})-{\bf m}^{(i)},{\bf x}({\bf r})} &=& h k_{-i} \prod_{j=1}^{\nu} 
N^{n^r_{i,j}} {\left(N - n^r_{i,j}\right)! \over N!}\ {x_j({\bf r})! \over 
\left(x_j({\bf r}) - n^r_{i,j}\right)!}\ . 
\end{eqnarray}
\nrtxt
The prefactors in (\ref{T_reac}) are chosen so that the mean-field kinetic 
equations can be written in a neat analytic form. The exclusion principle 
forbids those reactive events which produce particles of any species in excess 
of $N$. Therefore, we set:
\begin{eqnarray}
\label{exclusion}
T^R_{{\bf x}({\bf r}) + {\bf m}^{(i)},{\bf x}({\bf r})} = 0, & & 
\;\; \mbox{if $x_j + m^{(i)}_j > N$}, \nonumber \\
T^R_{{\bf x}({\bf r}) - {\bf m}^{(i)},{\bf x}({\bf r})} = 0, & &
\;\; \mbox{if $x_j - m^{(i)}_j > N$}
\end{eqnarray}
for any $i, j$. Substituting the transition probabilities (\ref{T_reac}), 
modified according to (\ref{exclusion}), into (\ref{E-op}) yields the matrix
elements of reaction evolution operator $\hat{W}^R$.

	Let us consider the mean-field dynamics of the reaction Markov chain. 
To begin, we note that in the mean-field limit (i.e. in the limit of a well-stirred 
system where the rate of diffusion is infinite), the particles are re-distributed 
among the nodes of the lattice instantaneously at every moment of time. Hence, 
the local correlations between different species are effectively destroyed at 
the same moment as the local reactive events create them. This implies that 
the particles populate every node independently of each other, and the local 
probability distribution is the time-dependent multivariate binomial:
\wdtxt
\begin{equation}
\label{binomial}
P({\bf x}({\bf r}),n) = P_B({\bf x}({\bf r}),n) = \prod_{j=1}^{\nu} 
{N \choose x_j({\bf r})} \left({\overline{x}_j(n) \over N} \right)^{x_j({\bf r})} 
\left(1 - {\overline{x}_j(n) \over N}\right)^{N - x_j({\bf r})}.
\end{equation}
\nrtxt
We have dropped the notation due to spatial coordinates in the average 
concentrations in (\ref{binomial}) since the system is homogeneous in the 
mean-field picture. It is easy to show that the kinetic equations in this 
limit have the following form:
\wdtxt
\begin{eqnarray}
\label{mfield}
\overline{x}_j(n+1) - \overline{x}_j(n) &=& \sum_{{\bf x}}^{} x_j 
\hat{W}^R P_B({\bf x},n) \nonumber \\
&=& h \sum_{i=1}^r \left( n^r_{i,j} - n^f_{i,j} 
\right) \left[k_i \prod_{k=1}^{\nu} \overline{x}_k^{n^f_{i,k}}(n) - k_{-i} 
\prod_{k=1}^{\nu} \overline{x}_k^{n^r_{i,k}}(n) \right] + 
Q_j(\overline{{\bf x}}(n)) \\
&=& K_j(\overline{{\bf x}}(n)). \nonumber 
\end{eqnarray}
\nrtxt			    
One observes that the first term on the second line of (\ref{mfield}) is 
the mass-action law (in the discrete-time notation) appropriate for the 
overall reaction mechanism. The additional term ${\bf Q}(\overline{{\bf x}}(n))$ 
describes the deviations of the mean-field dynamics from the mass-action law 
due to the exclusion corrections to the birth-death transition probabilities
(\ref{T_reac}). Eq. (\ref{mfield}) is the exact mean-field rate law for our 
model. In the application to Willamowski-R\"ossler reaction diffusion system 
in Sec. \ref{WR} we shall consider conditions where ${\bf Q}(\overline{{\bf x}}(n)$
is negligible and eq. (\ref{mfield}) reduces to the mass-action law.

	For a finite diffusion rate, neither (\ref{binomial}) nor (\ref{mfield})
will be valid. The nature of this breakdown of mean-field description forms
the central point of the present study. Note also that the binomial distribution 
(\ref{binomial}) depends on time only through its means $\overline{x}_j(n)$ 
which evolve according to (\ref{mfield}). We will use this time-dependent 
binomial distribution as the zeroth-order approximation to the full distribution 
in the following section.				    

\section{Perturbation Theory}
\label{theory}
							      
	Suppose that the system is not completely homogeneous, i.e. diffusion 
occurs at a finite rate determined by the value of diffusion coefficient $D$. 
Such a regime can be characterized by two time scales, $\tau_D$ and $\tau_R$, 
associated with diffusion and reaction processes, respectively. We
consider finite systems so that long wavelength diffusion modes are suppressed
and we may insure that $\tau_D \ll \tau_R$. In the context 
of our model $\tau_D$ can be the mean time required for a particle 
to travel the distance between two neighbouring nodes, and $\tau_R$ the inverse 
of the smallest eigenvalue of reaction evolution operator $\hat{W}^R$. 
The deviation from the mean-field limit is measured by the ratio ${\tau_D \over 
\tau_R}$, which will be denoted as $\gamma$ henceforth; $\gamma$ is equal to 
zero in the mean-field limit, and increases away from it.

	If $\gamma$ is small, the local probability distribution 
$P({\bf x}({\bf r}),t)$ obeys
\wdtxt
\begin{equation}
\label{RD_chain}
P({\bf x}({\bf r}),n+1) - P({\bf x}({\bf r}),n) = \left( \gamma 
\hat{W}^R + \hat{W}^D \right) P({\bf x}({\bf r}),n),
\end{equation}
\nrtxt
where $\hat{W}^R$ and $\hat{W}^D$ are the reaction and diffusion
evolution operators constructed in previous section.
We expect that the solution of (\ref{RD_chain}) monotonously approaches 
the local binomial distribution (\ref{binomial}) of the mean-field limit as 
$\gamma$ tends to zero. For a non-zero $\gamma$ this mean-field result is not
valid, and corrections to the local binomial distribution must be introduced.
In view of this, for small $\gamma$ reactions can be considered as a small 
perturbation to the pure diffusion process and perturbation theory can be 
applied to solve equation (\ref{RD_chain}). Thus, we write the local 
probability distribution in the form of a perturbation series: 
\wdtxt
\begin{equation}
\label{expansion}
P({\bf x}({\bf r}),n) = P_B({\bf x}({\bf r}),n) + 
\gamma P_1({\bf x}({\bf r}),n) + \gamma^2 P_2({\bf x}({\bf r}),n)
+ \dots,
\end{equation}
\nrtxt
with the local binomial distribution as the zeroth-order term.  Substituting 
(\ref{expansion}) into evolution equation (\ref{RD_chain}) gives:
\wdtxt
\begin{eqnarray}
\label{series_chain}
P_B({\bf x}({\bf r}),n+1) - P_B({\bf x}({\bf r}),n) &+&
\sum_{k=1}^{\infty} \gamma^k \{P_k({\bf x}({\bf r}),n+1) - 
P_k({\bf x}({\bf r}),n) \} \\
&=& \left(\gamma \hat{W}^R + \hat{W}^D \right) 
\left(P_B({\bf x}({\bf r}),n) + \sum_{k=1}^{\infty} \gamma^k 
P_k({\bf x}({\bf r}),n) \right). \nonumber
\end{eqnarray}
\nrtxt

	In order to extract the time dependence of each of the expansion terms from 
(\ref{series_chain}), we use a perturbation theoretic procedure similar to 
Chapman-Enskog development \cite{uhlenbeck,modce}. Specifically, we note that the 
first two terms in (\ref{series_chain}) are equivalent to a discrete derivative 
of the binomial distribution $P_B({\bf x}({\bf r}),n)$ with respect to time. 
Recalling that this distribution depends on time implicitly, through its means 
$\overline{{\bf x}}({\bf r},n)$, one can re-write the derivative as follows:
\wdtxt
\begin{equation}
\label{pb_deriv}
P_B({\bf x}({\bf r}),n+1) - P_B({\bf x}({\bf r}),n) = 
\Bigl[ \overline{{\bf x}}({\bf r},n+1) - \overline{{\bf x}}({\bf r},n) 
\Bigr] \; {\partial \over \partial \overline{{\bf x}}} P_B({\bf x}({\bf r}),n).
\end{equation}
\nrtxt

	The quantity $\overline{{\bf x}}({\bf r},n+1) - \overline{{\bf x}}
({\bf r},n)$ appearing in (\ref{pb_deriv}) is given by the kinetic equations
corresponding to (\ref{series_chain}). The latter can be obtained in the 
standard way, i.e. multiplying both sides of (\ref{series_chain}) by ${\bf x}
({\bf r})$ and summing over all possible combinations of the occupation numbers.
Using the properties of operators $\hat{W}^D$ and $\hat{W}^R$ 
introduced in Sec. \ref{model}, we find:
\wdtxt
\begin{equation}
\label{mean_ev}
\overline{{\bf x}}({\bf r},n+1) - \overline{{\bf x}}({\bf r},n) = 
D \: \Delta \overline{{\bf x}}({\bf r},n) + \gamma {\bf K}(\overline{{\bf x}}
({\bf r},n)) + 
\sum_{k=1}^{\infty} \gamma^{k+1} \: \left \langle {\bf x}
({\bf r}) \hat{W}^R P_k({\bf x}({\bf r}),n) \right \rangle, \nonumber
\end{equation}
\nrtxt
where ${\bf K}(\overline{{\bf x}}({\bf r},n))$ stands for mean-field dynamics 
(cf. eq. (\ref{mfield})), and
\begin{equation}
\label{angle_brackets}
\hspace{-12 pt}
\left \langle {\bf x}({\bf r}) \hat{W}^R P_k({\bf x}({\bf r}),n) \right 
\rangle = \sum_{{\bf x}({\bf r})} {\bf x}({\bf r}) \hat{W}^R P_k({\bf x}({\bf r}),n).
\end{equation}

	Before we proceed with the development of the solution to the evolution 
equation (\ref{series_chain}) one important question must be addressed, 
specifically that of the validity of the regular perturbation expansion 
(\ref{expansion}) for the case of far-from-equilibrium chemical systems.
It is well-known that the dynamics of such systems may be of relaxational
character, i.e. it exhibits well-defined fast and slow parts. This is manifested
by the significant jumps in reaction rates as well as the higher derivatives
of concentrations with respect to time which are observed as the system
switches from fast to slow dynamics, or vice versa. In principle, it is
quite possible that such jumps can cause a similar singular behaviour in 
the probability distribution $P({\bf x}({\bf r}),n)$, thereby making the
description in terms of a regular perturbation series invalid. In order to 
avoid such breakdown, we require that $\gamma$ be small enough for the 
following relations to hold at all ${\bf r}$ and $n$:
\begin{eqnarray}
\label{regularity}
P_i({\bf x}({\bf r}),n) &\ll& \gamma^{-1}, \\
{\left \langle {\bf x}({\bf r}) \hat{W}^R P_i({\bf x}({\bf r}),n) \right 
\rangle} &\ll& \gamma^{-1},\:\: i = 1,2,3,\dots \nonumber
\end{eqnarray}  
With this requirement satisfied, the regular perturbation method remains valid, 
at least on the level of description provided by equations (\ref{series_chain}) 
and (\ref{mean_ev}).
	
	Now, suppose that we want to determine the probability distribution up 
to the $k$-th order term. For that purpose, it is sufficient to retain in (\ref{mean_ev}) 
terms up to the order $\gamma^k$. Then, with the use of (\ref{mean_ev}) and (\ref{pb_deriv}), we 
can separate (\ref{series_chain}) into a set of $k$ equations, each describing 
evolution of one of the expansion terms $P_1({\bf x}({\bf r}),n), \dots, 
P_k({\bf x}({\bf r}),n)$. Thus, the equation for $P_1({\bf x}({\bf r}),n)$ 
reads:
\wdtxt
\begin{eqnarray}
\label{p1_ev}
\gamma P_1({\bf x}({\bf r}),n+1) &-& \gamma P_1({\bf x}({\bf r}),n) =
\left[ \hat{W}^D - \: D \: \Delta \overline{{\bf x}}({\bf r},n) \:
{\partial \over \partial \overline{{\bf x}}} \right] P_B({\bf x}({\bf r}),n) 
\\
&+& \gamma \left[ \hat{W}^R - \: {\bf K}(\overline{{\bf x}}({\bf r},n)) \: 
{\partial \over \partial \overline{{\bf x}}} \right] P_B({\bf x}({\bf r}),n) 
+ \gamma \hat{W}^D P_1({\bf x}({\bf r}),n), \nonumber
\end{eqnarray}
\nrtxt
and for $P_i({\bf x}({\bf r}),n),\: 1 < i \leq k$ (cancelling out the factor
$\gamma^i$):
\wdtxt
\begin{eqnarray}
\label{pi_ev}
P_i({\bf x}({\bf r}),n+1) - P_i({\bf x}({\bf r}),n) &=&
\hat{W}^R P_{i-1}({\bf x}({\bf r}),n) + \hat{W}^D P_i({\bf x}({\bf r}),n) \\
&-& \left \langle {\bf x}({\bf r}) \hat{W}^R P_i({\bf x}({\bf r}),n)
\right \rangle {\partial \over \partial \overline{{\bf x}}} 
P_B({\bf x}({\bf r}),n). \nonumber
\end{eqnarray}
\nrtxt
The solution of this system of inhomogeneous linear equations will 
be unique if we require:
\begin{eqnarray}
\label{solv_cond}
\Bigl \langle P_i({\bf x}({\bf r}),n) \Bigr \rangle &=& 0,\\
\Bigl \langle {\bf x}({\bf r}) P_i({\bf x}({\bf r}),n) \Bigr \rangle &=& 0, 
\:\: i = 1, \dots, k. \nonumber	  
\end{eqnarray}
Together with the solvability conditions (\ref{solv_cond}), (\ref{p1_ev}) 
and (\ref{pi_ev}) form a closed hierarchy from which one may evaluate the terms 
of the perturbation expansion (\ref{expansion}) to any desired order.

	In the remainder of this paper our attention will be limited to 
the first two terms of the series (\ref{expansion}). Therefore, we need only 
solve the lowest-order equation of the above hierarchy, i.e. (\ref{p1_ev}). 
By direct computation of $\hat{W}^D P_B({\bf x}({\bf r}),n)$ we may
show that the first term on the r.h.s. of (\ref{p1_ev}) is zero and, 
consequently, that $P_1({\bf x}({\bf r}),n+1) - P_1({\bf x}({\bf r}),n)$ 
does not contain contributions of order unity. 
For simplicity, we assume that the system consists of a single species. Using 
the definition of the diffusion evolution operator $\hat{W}^D$ (see 
(\ref{Diff_chain}) and (\ref{Wdiff})) we can write this term as:
\wdtxt
\begin{eqnarray}
\label{diff_pb}
& & {\chi({\bf r},n) \over N} \left[ \left(1 - {{x({\bf r})-1} \over N} \right) 
P_B(x({\bf r})-1,n) - \left(1 - {x({\bf r}) \over N} \right) P_B(x({\bf r}),n) 
\right] \\
&+& \left(1 - {\chi({\bf r},n) \over N} \right) \left[{{x({\bf r})+1} \over N}
P_B(x({\bf r})+1,n) - {x({\bf r}) \over N} P_B(x({\bf r}),n) \right] 
- {{\chi({\bf r},n) - \overline{x}({\bf r},n)} \over N} {\partial \over \partial 
\overline{x}} P_B(x({\bf r},n). \nonumber
\end{eqnarray}
\nrtxt
Taking advantage of two easily proven results,
\begin{eqnarray}
\label{corollary}
& &\left(1 - {{x-1} \over N} \right) P_B(x-1,n) = {x (N - \overline{x}(n)) 
\over N \overline{x}(n)} P_B(x,n), \nonumber \\
\\ 
& &{\partial \over \partial \overline{x}} P_B(x,n) = {x P_B(x,n) - 
(x+1) P_B(x+1,n) \over \overline{x}(n)}, \nonumber
\end{eqnarray}
we can show that (\ref{diff_pb}) vanishes. Generalization of this result
for a many-species system is straightforward due to the properties of the 
diffusion evolution operator (cf. (\ref{Diff_chain})).

	With the first term on the r.h.s. of (\ref{p1_ev}) eliminated, we can
obtain $P_1({\bf x}({\bf r}),n)$ by solving a simple initial-value problem.
Suppose that at $n=0$ the local probability distribution is purely binomial, 
i.e. $P_1({\bf x}({\bf r}),0) = 0$. Then, one finds from (\ref{p1_ev}):
\wdtxt
\begin{equation}
\label{p1}
P_1({\bf x}({\bf r}),n) = \sum_{n'=0}^{n-1} \left( \hat{I} + 
\hat{W}^D \right)^{n-n'-1} \left[ \hat{W}^R - {\bf K}(\overline{{\bf x}}
({\bf r}, n')) {\partial \over \partial \overline{{\bf x}}} \right] 
P_B({\bf x}({\bf r}), n'), 
\end{equation}
\nrtxt
where $\hat{I}$ is the identity operator ($I_{{\bf x}'{\bf x}} =
\delta_{{\bf x}'{\bf x}}$), and $\left( \hat{I} + \hat{W}^D \right)
^{n-n'-1}$ is to be understood as a time-ordered product of $n-n'-1$ terms 
$\left( \hat{I} + \hat{W}^D \right)$, with the leftmost term taken 
at the moment of time $n-1$ and the rightmost at $n'+1$. In Appendix A we 
prove that (\ref{p1}) does satisfy the solvability conditions (\ref{solv_cond}).

	The quantities whose dynamics will be of interest to us are the 
global concentrations $\overline{{\bf x}}(t)$, i.e. the local concentrations 
$\overline{{\bf x}}({\bf r},t)$ averaged over all nodes of the lattice. 
(Henceforth, whenever we omit ${\bf r}$ in the arguments of the concentrations,
the global concentrations are meant.) The kinetic equations for the local 
concentrations are given by (\ref{mean_ev}); at the level of present approximation 
we can truncate the infinite series on the r.h.s. of (\ref{mean_ev}) after the 
term of order $\gamma^2$. To obtain kinetic equations for the global concentrations 
we substitute $P_1({\bf x}({\bf r}),n)$ from (\ref{p1}) into (\ref{mean_ev}), 
sum both sides over all ${\bf r}$, and divide by the total number of nodes $M$. 
Since diffusion conserves the total number of particles in the system, the 
pure diffusion term in (\ref{mean_ev}) yields zero on summation and we obtain 
for the global concentration of the $i$-th species:
\wdtxt
\begin{equation}
\label{global_ev}
\overline{x}_i(n+1) - \overline{x}_i(n) = {\gamma \over M} \sum_{{\bf r}}^{} K_i(
\overline{{\bf x}}({\bf r},n)) + {\gamma^2 \over M} \sum_{{\bf r}}^{} 
\sum_{n'=0}^{n-1} \left \langle x_i({\bf r}) \hat{W}^R \left( \hat{I} + 
\hat{W}^D \right)^{n-n'-1} \hat{S} P_B({\bf x}({\bf r}), n') \right \rangle,
\end{equation}
\nrtxt
where $\hat{S} = \hat{W}^R - {\bf K}(\overline{{\bf x}}({\bf r},n')) {\partial 
\over \partial \overline{{\bf x}}}$, and $\overline{{\bf x}}(n)$ is the vector 
of global concentrations.

	If diffusion is sufficiently fast ($\gamma \ll 1$), the system remains 
nearly homogeneous at all times, i.e. $\overline{x}_i({\bf r},n)$ are slowly 
varying functions of ${\bf r}$ which do not deviate significantly from their 
spatial averages $\overline{x}_i(n)$. Hence, if we put:
\begin{equation}
\label{inhomog}
\overline{x}_i({\bf r},n) \approx \overline{x}_i(n)
\end{equation}
for all ${\bf r}$ and $n$, the summation over ${\bf r}$ in (\ref{global_ev}) 
becomes trivial, and we find:
\wdtxt
\begin{equation}
\label{global_ev_app}
\overline{x}_i(n+1) - \overline{x}_i(n) = \gamma K_i(\overline{{\bf x}}(n)) + 
\gamma^2 \sum_{n'=0}^{n-1} \left \langle x_i({\bf r}) \hat{W}^R \left(\hat{I} 
+ \hat{W}^D \right)^{n-n'-1} \hat{S} P_B({\bf x}, n') \right \rangle,
\end{equation}
\nrtxt
where the local concentrations appearing implicitly in $\hat{W}^D$,
$\hat{S}$, and $P_B({\bf x},n)$ in the last term are replaced by global 
concentrations $\overline{{\bf x}}(n)$. 

	The kinetic equations (\ref{global_ev_app}) do not readily offer a deeper
insight into dynamics of the model since the memory function in the last
term cannot be easily evaluated for general reaction mechanisms. However, as 
we will show in the following section, for specific reaction mechanisms it is
possible to write these equations in a tidy analytic form which allows clear
physical interpretation.

\section{Application to Willamowski-R\"ossler model}
\label{WR}    

	We now apply the above formalism to a reaction-diffusion system 
with Willamowski-R\"ossler reaction mechanism \cite{wr}. This mechanism 
consists of the following elementary steps:
\begin{eqnarray}
A_1 +X &\mathrel{
\mathop{\kern0pt {\rightleftharpoons}}\limits^{{k_1}}_{k_{-1}}}& 2X,\;\;
X+Y
\mathrel{\mathop{\kern0pt {\rightleftharpoons}}\limits^{{k_2}}_{k_{-2}}}
2Y,\nonumber \\
A_5 +Y &
\mathrel{\mathop{\kern0pt {\rightleftharpoons}}\limits^{{k_3}}_{k_{-3}}}&
A_2,\;\;
X+Z
\mathrel{\mathop{\kern0pt {\rightleftharpoons}}\limits^{{k_4}}_{k_{-4}}}
A_3, \nonumber \\
A_4 +Z &
\mathrel{\mathop{\kern0pt {\rightleftharpoons}}\limits^{{k_5}}_{k_{-5}}}&
2Z\;. \nonumber
\end{eqnarray}
Here $A_1$, $A_4$, $A_5$ are initiators, and $A_2$, $A_3$ are final 
products; concentrations of all these species are held constant by external 
fluxes. The intermediates whose dynamics is followed are $X$, $Y$, and $Z$. 
The mass-action law for this model has the following form (indices 1,2,3 refer
to $X$, $Y$, and $Z$ species, respectively):
\wdtxt
\begin{eqnarray}
\label{WR_mass_ac}
{d \overline{x}_1 \over d t} &=& \kappa_1 \overline{x}_1(t) - \kappa_{-1} \overline{x}_1^2(t) - 
\kappa_2 \overline{x}_1(t) \overline{x}_2(t) + \kappa_{-2} \overline{x}_2^2(t) - \kappa_4 \overline{x}_1(t) 
\overline{x}_3(t) + \kappa_{-4} = R_1(\overline{x}_1(t),\overline{x}_2(t),\overline{x}_3(t))\;, \nonumber \\     
{d \overline{x}_2 \over d t} &=& \kappa_2 \overline{x}_1(t) \overline{x}_2(t) - \kappa_{-2} \overline{x}_2^2(t) 
- \kappa_{3} \overline{x}_2(t) + \kappa_{-3} = R_2(\overline{x}_1(t),\overline{x}_2(t))\;, \\
{d \overline{x}_3 \over d t} &=& - \kappa_4 \overline{x}_1(t) \overline{x}_3(t) + \kappa_5 \overline{x}_3(t) 
- \kappa_{-5} \overline{x}_3^2(t) + \kappa_{-4} = R_3(\overline{x}_1(t),\overline{x}_3(t))\;, 
\nonumber
\end{eqnarray}
\nrtxt
where the concentrations of initiators and products are incorporated in 
the values of rate constants $\kappa_i$, \mbox{$i = 1,\dots,5$}. Equations 
(\ref{WR_mass_ac}) have been considered in earlier studies and are known to
give rise to a period-doubling cascade leading to chaotic attractor \cite{wu,scott,aguda}.	     

	We begin our analysis by calculating the contributions that the reactive 
steps of Willamowski-R\"ossler mechanism make to the memory function in 
(\ref{global_ev_app}). Below, the contributions from mono- and bimolecular 
steps are considered in turn.

\subsection{Monomolecular steps}
\label{monomolecular}

	All monomolecular steps in Willamowski-R\"ossler model have the 
following form:
$$	
A_i \mathrel{\mathop{\kern 0pt {\rightarrow}}\limits^{k}_{}} 
n_j X_j + n_l X_l + n_m A_m, \;\; n_j, n_l, n_m = 0,1.
$$					  
The matrix elements of the reaction evolution operator corresponding to 
such steps can be obtained from the general expression for reactive transition 
probabilities (\ref{T_reac}). We find: 
\[
W^R_{{\bf x}{\bf x}'} = \left \{
\begin{array}{ll}
- h \kappa \left( 1 - m_j \delta_{x_j,N} \right) \left( 1 - m_l \delta_{x_l,N} \right), 
& \mbox{if ${\bf x}' = {\bf x}$,} \\	
h \kappa, \;\; \mbox{if ${\bf x}' = {\bf x} - {\bf m}$,} & \\
0, \;\; \mbox{for all other ${\bf x}'$,} &
\end{array} \right.
\]		  
where the Kronecker deltas account for the exclusion principle, the concentration 
of species $A_i$ is incorporated in the value of $\kappa$ and vector ${\bf m}$ 
is defined through the stoichiometric coefficients as outlined in Sec. 
\ref{model}.

	At this point, in order to facilitate further calculations, we neglect 
the terms in the memory function that arise from the exclusion principle. Such
approximation is justified for Willamowski-R\"ossler model, since those terms
can always be made arbitrarily small by a simple rescaling of concentrations.
As an example, suppose that $N = 10$ and the concentrations are scaled so that
they practically never exceed $1.5$ particles per node. Given that the reactive
transition probabilities are of the order of $10^{-3}$ -- $10^{-4}$, numerical
estimates show that the terms due to the exclusion principle are $2$ -- $3$ 
orders of magnitude smaller than all the other terms in (\ref{global_ev_app}).
(The concentration scaling and the value of N used in the lattice gas automaton 
simulations of Willamowski-R\"ossler model below are the same as considered 
here.)

	The contribution to the memory function can now be evaluated in a fairly
straightforward way. We substitute the evolution operator from above in place 
of the leftmost $\hat{W}^R$ in the last term of (\ref{global_ev_app}), and 
calculate the average. After a few transformations, we arrive at:
\begin{equation}
\label{mono_contr}
\gamma^2 h \kappa \sum_{n'=0}^{n-1} \left \langle \left( \hat{I} + \hat{W}^D 
\right)^{n-n'-1} \hat{S} P_B({\bf x}, n') \right \rangle. 
\end{equation}
One immediately recognizes in (\ref{mono_contr}) one of the two quantities 
(to within a constant multiplier) that we have dealt with in a different context 
in Appendix A (cf. (\ref{p1_1_solv})). Using the results obtained in Appendix 
A we find that (\ref{mono_contr}) vanishes. Therefore, we conclude that 
monomolecular reactive steps in Willamowski-R\"ossler mechanism do not contribute 
to the memory function in (\ref{global_ev_app}).

\subsection{Bimolecular steps}
\label{bimolecular}
	The bimolecular steps in Willamowski-R\"ossler model fall into three 
classes; these are listed below, along with the corresponding matrix elements 
of the $\hat{W}^R$ operator: 
$$
\hspace{-22 pt}
1) \;\; 2 X_i \mathrel{\mathop{\kern 0pt {\rightarrow}}\limits^{k}_{}} 
X_i + n_j X_j + n_l A_l, \;\; n_j,n_l = 0, 1,
$$
\[
\hspace{-16 pt}
W^R_{{\bf x}{\bf x}'} = \left \{ 
\begin{array}{ll} 
- {N \over {N-1}} h k x_i \left( x_i-1 \right) \left( 1 - m_j \delta_{x_j, N} 
\right), & \mbox{if ${\bf x}' = {\bf x}$}, \\
{N \over {N-1}} h k \left( x_i - m_i \right) \left( x_i - m_i - 1 
\right),\;\; \mbox{if}& {\bf x}' = {\bf x} - {\bf m}, \\
0, \;\; \mbox{for all other ${\bf x}'$}, &
\end{array} \right.
\]	 
$$
2) \;\; X_i + X_j \mathrel{\mathop{\kern 0pt {\rightarrow}}\limits^{k}_{}}
n_i X_i + n_j A_j, \;\; n_i = 0,2; n_j = 0,1,
$$				   
\[
W^R_{{\bf x}{\bf x}'} = \left \{ 
\begin{array}{ll} 
- h k x_i x_j \left( 1 - {m_i+1 \over 2} \delta_{x_i,N} \right), & 
\mbox{if ${\bf x}' = {\bf x}$}, \\
  h k \left( x_i - m_i \right) \left( x_j - m_j \right),& 
\mbox{if ${\bf x}' = {\bf x} - {\bf m}$}, \\
  0,     & \mbox{for all other ${\bf x}'$},
\end{array} \right.
\]
$$	   
\hspace{-124 pt}
3) \;\; A_i + X_j \mathrel{\mathop{\kern 0pt {\rightarrow}}\limits^{k}_{}}
2 X_j,
$$
\[
W^R_{{\bf x}{\bf x}'} = \left \{ 
\begin{array}{ll} 
- h \kappa x_j \left( 1 - \delta_{x_j,N} \right), & 
\mbox{if ${\bf x}' = {\bf x}$}, \\
  h \kappa \left( x_j - m_j \right),& 
\mbox{if ${\bf x}' = {\bf x} - {\bf m}$}, \\
  0,     & \mbox{for all other ${\bf x}'$},
\end{array} \right.
\]		   
where $m_j = 1$ and the concentration of $A_i$ is incorporated into the value 
of $\kappa$.

	We neglect the terms arising from the exclusion principle on the basis 
of the rescaling argument given above, and calculate the contributions to the 
memory function due to these reactive steps. We find: 
\wdtxt
\begin{eqnarray}
\label{bi_contr}
&1)& \;\;\; {N \over {N-1}} \; \gamma^2 h k \sum_{n'=0}^{n-1} \left \langle 
x_i \left( x_i-1 \right) \left( \hat{I} + \hat{W}^D \right)^
{n-n'-1} \hat{S} P_B({\bf x}, n') \right \rangle, \nonumber \\
&2)& \;\;\; \gamma^2 h k \sum_{n'=0}^{n-1} \left \langle x_i 
x_j \left( \hat{I} + \hat{W}^D \right)^{n-n'-1} \hat{S} 
P_B({\bf x}, n') \right \rangle. \\
&3)& \;\;\; \gamma^2 h \kappa \sum_{n'=0}^{n-1} \left \langle x_j 
\left( \hat{I} + \hat{W}^D \right)^{n-n'-1} \hat{S} 
P_B({\bf x}, n') \right \rangle. \nonumber
\end{eqnarray}
\nrtxt
Note that the last contribution in (\ref{bi_contr}) is identical (to
within a constant multiplier) to one of the quantities considered in Appendix 
A (cf. (\ref{p1_2_solv})). From the results of Appendix A we infer that this 
contribution vanishes. Also, using the method of Appendix A one can show that 
the first two contributions in (\ref{bi_contr}) can be reduced to the following 
memory terms with a simple exponential kernel:
\wdtxt
\begin{eqnarray}
\label{exp_kernel}
&1)& \;\; {N \over {N-1}} \; \gamma^2 h k \sum_{n'=0}^{n-1} \left( 1 - {2 \over 
\nu N} \right)^{n-n'-1} \left \langle x^2_i \hat{S} P_B({\bf x}, n' h) 
\right \rangle, \nonumber \\
&2)& \;\; \gamma^2 h k \sum_{n'=0}^{n-1} \left( 1 - {2 \over \nu N} 
\right)^{n-n'-1} \left \langle x_i x_j \hat{S} P_B({\bf x}, n' h) 
\right \rangle.
\end{eqnarray}				    
\nrtxt

\vspace{-14 pt}
	Summarizing the results obtained above, we conclude that the only 
reactive steps that contribute to the memory function in (\ref{global_ev_app})
are those involving reactions between the particles of intermediate species.
The contribution is given by the first term in (\ref{exp_kernel}) if the reacting 
particles are of the same chemical nature, and by the second term otherwise.
	
	By inspection of the mechanism, we find four contributing steps for 
the $X$ species:
\vspace{-2 pt}
\[
\begin{array}{rcl}
2 X & 
\mathrel{\mathop{\kern0pt {\rightarrow}}\limits^{{k_{-1}}}_{}}& A_1 + X, \\
2 Y 
&\mathrel{\mathop{\kern0pt {\rightarrow}}\limits^{{k_{-2}}}_{}}& X + Y, \\
X+Y
&\mathrel{\mathop{\kern0pt {\rightarrow}}\limits^{{k_2}}_{}}& 2Y, \\
X+Z
&\mathrel{\mathop{\kern0pt {\rightarrow}}\limits^{{k_4}}_{}}& A_3, 
\end{array}
\]		   	
two for the $Y$ species:
\[
\begin{array}{rcl}
2 Y 
&\mathrel{\mathop{\kern0pt {\rightarrow}}\limits^{{k_{-2}}}_{}}& X + Y, \\
X+Y
&\mathrel{\mathop{\kern0pt {\rightarrow}}\limits^{{k_2}}_{}}& 2Y,
\end{array}
\]
and two for the $Z$ species:
\[
\begin{array}{rcl}
2 Z & 
\mathrel{\mathop{\kern0pt {\rightarrow}}\limits^{{k_{-5}}}_{}}& A_4 + Z, \\ 
X+Z
&\mathrel{\mathop{\kern0pt {\rightarrow}}\limits^{{k_4}}_{}}& A_3. 
\end{array}
\]
To obtain the kinetic equations for the global concentrations of each species, 
one simply adds contributions appropriate for the steps in each of the three 
groups above and substitutes the result in place of the memory term in 
(\ref{global_ev_app}). Proceeding in this way, we arrive at:
\wdtxt
\begin{eqnarray}
\label{WR_global_ev_a}				   
\overline{x}_1(n+1) - \overline{x}_1(n) = \gamma h R_1(\overline{{\bf x}}(n)) 
&-& \gamma^2 h {N \over {N-1}} \sum_{n'=0}^{n-1} \left( 1 - \lambda \right)^
{n-n'-1} \Bigl[ \kappa_{-1} C_{x_1,x_1}(n') - \kappa_{-2} C_{x_2,x_2}(n') 
\Bigr] \\ 
&-& \gamma^2 h \sum_{n'=0}^{n-1} \left( 1 - \lambda \right)^{n-n'-1} \Bigl[ 
\kappa_2 C_{x_1,x_2}(n') + \kappa_4 C_{x_1,x_3}(n') \Bigr], \nonumber \\
\nonumber \\
\label{WR_global_ev_b}				   
\overline{x}_2(n+1) - \overline{x}_2(n) = \gamma h R_2(\overline{{\bf x}}(n)) 
&-& \gamma^2 h \sum_{n'=0}^{n-1} \left( 1 - \lambda \right)^{n-n'-1} \left[
{N \kappa_{-2} \over {N-1}} C_{x_2,x_2}(n') - \kappa_{2} C_{x_1,x_2}(n') 
\right], \\ 
\nonumber \\
\label{WR_global_ev_c}				   
\overline{x}_3(n+1) - \overline{x}_3(n) = \gamma h R_3(\overline{{\bf x}}(n)) 
&-& \gamma^2 h  \sum_{n'=0}^{n-1} \left( 1 - \lambda \right)^{n-n'-1} \left[ 
{N \kappa_{-5} \over {N-1}} C_{x_3,x_3}(n') + \kappa_4 C_{x_1,x_3}(n') 
\right],
\end{eqnarray}
\nrtxt	      
where
\begin{eqnarray} 
\label{c_def}
&C&_{x_i,x_j}(t) = \left \langle x_i x_j \hat{S} P_B({\bf x},n) \right 
\rangle, \nonumber \\
&\hat{S}& = \hat{W}^R - h {\bf R}(\overline{{\bf x}}(n)) \; 
{\partial \over \partial \overline{{\bf x}}},
\end{eqnarray}
and $R_i(\overline{{\bf x}}(n)), \; i = 1,2,3$ are the mass-action law terms from (\ref{WR_mass_ac}),
$\lambda = {2 \over \nu N} = {2 \over 3 N}$ for Willamowski-R\"ossler model. 
The procedure which we used above to evaluate the memory terms in (\ref{global_ev_app}) 
by neglecting terms due to the exclusion principle can be extended to any 
reaction mechanism involving only mono- and bimolecular steps. Note that in 
the scaling limit where the terms due to the exclusion principle can be omitted 
in evaluating all average quantities, the mean-field reaction dynamics is entirely 
due to the mass-action law, i.e. ${\bf K}(\overline{{\bf x}}(n)) = h {\bf R}(\overline{{\bf x}}(n))$ 
(cf. (\ref{mfield})). The continuous-time form of eqns (\ref{WR_global_ev_a}) -- (\ref{WR_global_ev_c}) 
is derived in Appendix B.
	
	In order to evaluate the averages appearing in the equations 
(\ref{WR_global_ev_a}) -- (\ref{WR_global_ev_c}) one needs to know the
matrix elements of the full reaction evolution operator $\hat{W}^R$. The latter
are defined through the reactive transition probabilities which, for Willamowski-R\"ossler 
model, are as follows:
\wdtxt
\begin{eqnarray}	       
\label{reac_prob_WR}
&T&^R_{x_1+1,x_2,x_3;{\bf x}} = h \kappa_1 x_1 \left( 1 - \delta_{x_1,N} 
\right), \hskip 1.15 cm
T^R_{x_1-1,x_2,x_3;{\bf x}} = {N \over {N-1}} h \kappa_{-1} x_1 \left( 
x_1 - 1 \right), \nonumber \\
&T&^R_{x_1-1,x_2+1,x_3;{\bf x}} = h \kappa_2 x_1 x_2 \left( 1 - \delta_
{x_2,N} \right), \;\;\;\;
T^R_{x_1+1,x_2-1,x_3;{\bf x}} = {N \over {N-1}} h \kappa_{-2} x_2 \left( 
x_2 - 1 \right) \left( 1 - \delta_{x_1,N} \right), \nonumber \\
&T&^R_{x_1,x_2-1,x_3;{\bf x}} = h \kappa_3 x_2, 
\hskip 2.93 cm 
T^R_{x_1,x_2+1,x_3;{\bf x}} = h \kappa_{-3} \left(1 - \delta_{x_2,N} 
\right), \\
&T&^R_{x_1-1,x_2,x_3-1;{\bf x}} = h \kappa_4 x_1 x_3, \hskip 2.2 cm
T^R_{x_1+1,x_2,x_3+1;{\bf x}} = h \kappa_{-4} \left( 1 - \delta_{x_1,N} 
\right) \left( 1 - \delta_{x_3,N} \right),
\nonumber \\
&T&^R_{x_1,x_2,x_3+1;{\bf x}} = h \kappa_5 x_3 \left( 1 - \delta_{x_3,N} 
\right), \hskip 1.15 cm
T^R_{x_1,x_2,x_3-1;{\bf x}} = {N \over {N-1}} h \kappa_{-5} x_3 \left( 
x_3 - 1 \right). \nonumber
\end{eqnarray}
\nrtxt
Here ${\bf x} \equiv (x_1,x_2,x_3)$. Using the transition probabilities (\ref{reac_prob_WR}) to compute 
$C_{x_i,x_j}$, we obtain:
\pagebreak
\wdtxt		
\begin{eqnarray}
\label{WR_mem_func}
C_{x_1,x_1}(t) &=& 2 h \left[ \kappa_1 \overline{x}_1(t) - \kappa_{-1} \overline{x}^2_1(t) 
\left( 1 - {\overline{x}_1(t) \over N} \right) + {\kappa_{-2} \over N} \overline{x}_1(t)\overline{x}^2_2(t) 
+ {\kappa_{-4} \over N} \overline{x}_1(t) \right], \nonumber \\
C_{x_2,x_2}(t) &=& 2 h \left[ \kappa_2 \overline{x}_1(t) \overline{x}_2(t) - \kappa_{-2} \overline{x}^2_2(t) 
\left( 1 - {\overline{x}_2(t) \over N} \right) + {\kappa_{-3} \over N} \overline{x}_2(t) \right], 
\nonumber \\
C_{x_3,x_3}(t) &=& 2 h \left[ \kappa_5 \overline{x}_3(t) - \kappa_{-5} \overline{x}^2_3(t) 
\left( 1 - {\overline{x}_3(t) \over N} \right) + {\kappa_{-4} \over N} \overline{x}_3(t) \right], 
\\ 
C_{x_1,x_2}(t) &=& h \left[ \kappa_{-2} \overline{x}^2_2(t) \left(1 - {2 \overline{x}_2(t) \over N} 
\right) - \kappa_2 \overline{x}_1(t) \overline{x}_2(t) \left( 1 - {{\overline{x}_2(t) -\overline{x}_1(t)} 
\over N} \right) \right], \nonumber \\
C_{x_1,x_3}(t) &=& h \left[ \kappa_{-4} - \kappa_4 \overline{x}_1(t) \overline{x}_3(t) \left( 1 - {{
\overline{x}_1(t) + \overline{x}_3(t)} \over N} \right) \right]. \nonumber
\end{eqnarray}
\nrtxt
Equations (\ref{WR_global_ev_a}) -- (\ref{WR_global_ev_c}) together with 
(\ref{WR_mem_func}) describe the dynamics of the global concentrations within 
terms of order $\gamma^2$ for Willamowski-R\"ossler reaction-diffusion system. 						       

	We conclude this discussion by proving that the dynamics predicted by 
eqns (\ref{WR_global_ev_a}) -- (\ref{WR_global_ev_c}) is consistent with the 
initial assumptions of the theory, i.e. that it converges to mass-action law 
dynamics in the well-stirred regime (i.e. $\gamma \rightarrow 0$). Because $h$ 
is arbitrary, we observe that assigning a smaller value to $\gamma$ simply 
amounts to following the dynamics due to (\ref{WR_global_ev_a}) -- (\ref{WR_global_ev_c}) on a smaller 
time scale. This implies that under a rescaling of time 
$h' = \gamma h$, the limit $\gamma \rightarrow 0$ is formally equivalent to 
the limit $h' \rightarrow 0$ with $\gamma$ fixed. Defining a new discrete 
(real-valued) time variable $t_n = {n \over \gamma} h'$, we can re-cast 
(\ref{WR_global_ev_a}) -- (\ref{WR_global_ev_c}) in the following form (we put 
$\gamma = 1$ and consider only eqn. (\ref{WR_global_ev_b}), 
for simplicity):
\wdtxt
\begin{equation}
\label{resc_time}
\overline{x}_2(t_n + h') - \overline{x}_2(t_n) = h' R_2(\overline{{\bf x}}(t_n))
- {h'}^2 \sum_{n'=0}^{n-1} \left( 1 - \lambda \right)^{{t_n-t_{n'}-h' \over h'}}  
\left[{N \over {N-1}} {\kappa_{-2} C_{x_2,x_2}(t_{n'}) \over h'} - {\kappa_{2} 
C_{x_1,x_2}(t_{n'}) \over h'} \right]. 
\end{equation}					
\nrtxt
Keeping $t_n$ and $t_{n'}$ constant as we let $h'$ tend to zero, we obtain:
\wdtxt
\begin{eqnarray}
\label{mfield_limit}
{d \overline{x}_2(t) \over d t} &=& R_2(\overline{{\bf x}}(t)) 
- \lim_{h' \rightarrow 0} \int_0^{t-h'} e^{- {\lambda (t-t'-h') \over h'}}  
\left[{N \over {N-1}} {\kappa_{-2} C_{x_2,x_2}(t') \over h'} - {\kappa_{2} 
C_{x_1,x_2}(t') \over h'} \right] d t' \nonumber \\ 
&=& R_2(\overline{{\bf x}}(t)).
\end{eqnarray}
\nrtxt
The result on the last line follows since the kernel of the integral
term vanishes exponentially for all $t - t'$. Similar results hold for 
eqns (\ref{WR_global_ev_a}) and (\ref{WR_global_ev_c}) as well.

\subsection{Numerical results}
\label{numerical}  

\begin{figure}[htbp]
\begin{center}
\leavevmode
\epsffile{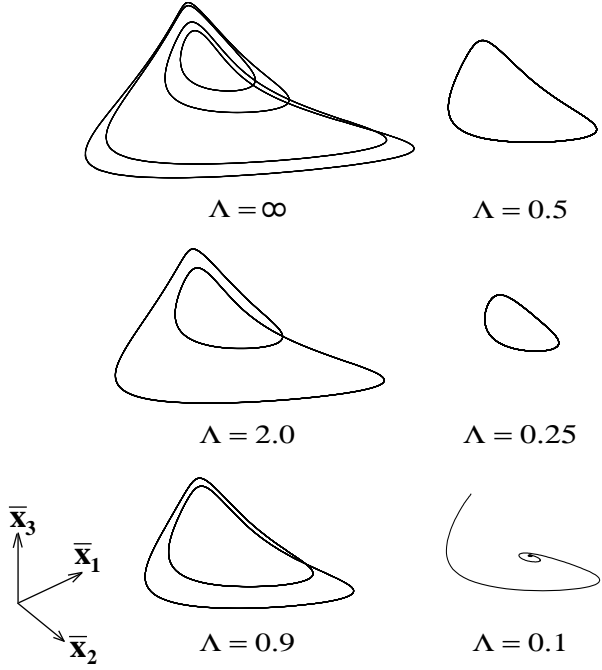}
\end{center}
\caption{Phase space trajectories obtained from equations (\protect 
\ref{WR_global_ev_a}) -- (\protect \ref{WR_global_ev_c}) for different values 
of $\Lambda$. Mass-action law trajectory ($\Lambda = \infty$) is period-4. 
Indices $1$, $2$, and $3$ in the coordinate labels correspond to $X$, $Y$, 
and $Z$ species, respectively. All trajectories are drawn to the same scale.
The values of $\Lambda$ are given in units of~${1 \over h}$ ($h = 5 \cdot 10^{-4}$).}  
\label{fig1}
\end{figure}

	We have solved numerically the continuous-time form of the equations 
(\ref{WR_global_ev_a}) -- (\ref{WR_global_ev_c}) (cf. Appendix B) using Euler's 
algorithm; the integral terms were evaluated at every time step by Simpson's 
method. In Fig.~\ref{fig1} we present the phase space trajectories obtained 
for different values of the diffusion coefficient (as measured by the exponent of 
the memory kernel $\Lambda$) as well as for the mean-field regime ($\Lambda = 
\infty$). The values of rate constants used in these calculations are as follows: 
$\kappa_1 = 31.2, \kappa_{-1} = 0.2, \kappa_2 = 1.533, \kappa_{-2} = 0.1, 
\kappa_3 = 10.8, \kappa_{-3} = 0.12, \kappa_4 = 1.02, \kappa_{-4} = 0.01, 
\kappa_5 = 16.5, \kappa_{-5} = 0.5$; this choice corresponds to the period-4 
regime in the mass-action dynamics. The integration time step $h$ was equal to
$5 \cdot 10^{-4}$.
\vspace*{24 pt}

	It is evident from Fig.~\ref{fig1} that, as the diffusion becomes slower 
(i.e., the value of $\Lambda$ becomes smaller), the trajectories are monotonously
transformed so that they traverse the period-doubling cascade of the mass-action 
law (\ref{WR_mass_ac}) in the backward direction: the smaller the value of the 
diffusion coefficient (with rate constants unchanged), the less the dynamics 
of the global concentrations corresponds to the mass-action law. 
\vspace*{\fill}
\eject	 
Furthermore, 
this breakdown of mass-action description can be modeled by a parametric shift 
in the mass-action law (along with rescaling of time and, possibly, concentrations) 
such that the dynamics proceeds from the period-4 regime (corresponding to the 
mean-field regime in Fig.~\ref{fig1}) to period-2, period-1, and finally to stable 
focus. 			       

	We now turn to the comparison of the numerical solutions of equations
(\ref{WR_global_ev_a}) -- (\ref{WR_global_ev_c}) and the dynamics observed in
the lattice gas automaton simulations of Willamowski-R\"ossler model. The
lattice gas automaton used in these simulations was implemented in essentially 
the same way as in the earlier studies \cite{wu,LGA}; the only difference is
in the nature of the diffusion rule, which we modify so that it corresponds
to the diffusion evolution operator (\ref{Wdiff}). The simulations were 
performed for a triangular lattice of size $200 \times 200$ nodes, with time 
step $h = 5 \cdot 10^{-4}$. The exclusion parameter $N$ was equal to $10$ (i.e.
no more than $10$ particles of each species were allowed to occupy any node at 
any time). All concentrations were scaled by a factor of $40$ so that they do 
not exceed $1.5$ particles per node throughout the simulations. 

	In a lattice gas automaton simulation, the local probability distribution 
evolves in time according to the following equation:
\vspace{36 pt}
\wdtxt
\begin{equation}
\label{LGA_ev}
P({\bf x}({\bf r}), n+1) = \left( \hat{I} + \hat{W}^R \right)^
{m_R} \left( \hat{I} + \hat{W}^D \right)^{m_D} P({\bf x}({\bf r}), n),
\end{equation}
\nrtxt
where $m_R$, $m_D$ are positive integers, and $\hat{I}$ is the identity
operator. Eq. (\ref{LGA_ev}) is approximately equivalent to the reaction-diffusion 
evolution equation (\ref{RD_chain}), with parameter $\gamma$ given,
to the leading order, by the ratio ${m_R \over m_D}$. This fact, along with
definition of the parameter $\Lambda$ ($\Lambda = {2 \over 3 N \gamma h}$ for 
Willamowski-R\"ossler model), allows us to approximately relate the results of 
an automaton simulation with a certain value of ${m_R \over m_D}$ to a solution 
of the kinetic equations (\ref{WR_global_ev_a}) -- (\ref{WR_global_ev_c}) with 
an appropriate value of $\Lambda$.

\begin{figure}[htbp]
\begin{center}
\leavevmode
\epsffile{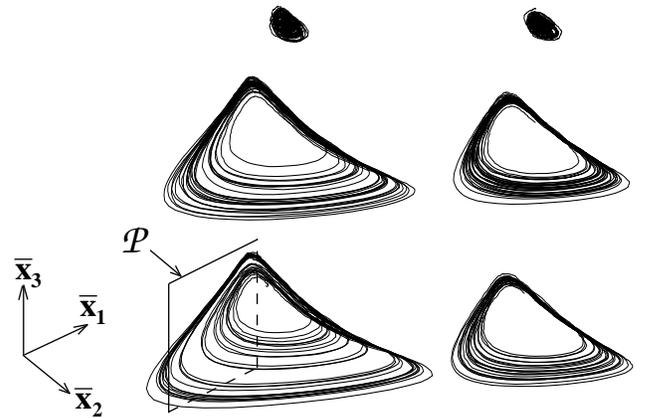}
\end{center}
\caption{Phase space trajectories from lattice gas automaton simulations for 
$\kappa_2 = 1.5$ (left) and $\kappa_2 = 1.4$ (right). The mass-action law 
dynamics is period-2 and period-1, respectively. For each parametric regime,
the value of $\Lambda$ (in units of~${1 \over h}$) are: $0.667$ (top panel), 
$13.333$ (middle panel), and $\infty$, (bottom panel). All trajectories are 
drawn to the same scale. Concentration scaling parameter is $40$.}  
\label{fig2}
\end{figure}

	In Fig.~\ref{fig2} we present phase-space trajectories from typical 
lattice gas automaton simulations for the parametric regimes where the
mass-action law dynamics is period-1 and period-2. All rate constants except 
$\kappa_2$ are the same as in Fig.~\ref{fig1}; $\kappa_2$ is equal to $1.4$ 
and $1.5$, respectively. For each regime, $\Lambda$ was equal to $0.667$, 
$13.333$, and $\infty$, the latter corresponding to the limit of well-stirred 
system. The simulations of well-stirred system were performed by re-seeding 
the nodes of the lattice according to binomial distribution at every time step.

	We observe that the trajectories from these simulations exhibit the
same general trend as those computed using kinetic equations 
(\ref{WR_global_ev_a}) -- (\ref{WR_global_ev_c}) in Fig.~\ref{fig1}. 
Namely,
note that for the regime where the mass-action law dynamics is period-2,
the trajectory is gradually transformed from noisy period-2 ($\Lambda = \infty$)
to noisy period-1 ($\Lambda = 13.333$) and finally to the stable focus 
($\Lambda = 0.667$). In order to investigate this effect in more detail, we 
construct a Poincar\'e section for the trajectories from automaton simulation. 
The half-plane of section (denoted by ${\cal P}$) is shown in Fig.~\ref{fig2}, 
bottom left panel, and defined as follows: 
$
\{ {\cal P}: \overline{x}_1 \geq 0.3; \overline{x}_2 = 0.25; \overline{x}_3 \geq 0 \}.
$

\begin{figure}[htbp]
\begin{center}
\leavevmode
\epsffile{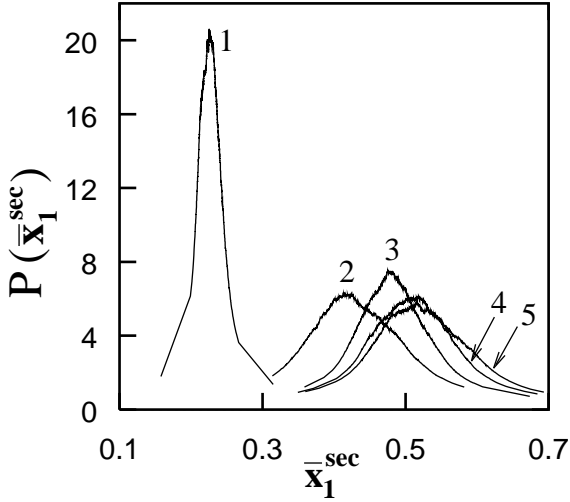}
\end{center}
\caption{Distribution of concentration of $X$ species on Poincare section \protect
${\cal P}$ for different values of $\Lambda$. The mass-action law dynamics is 
period-1 ($\kappa_2 = 1.4$). The graphs are numbered according to the value of 
$\Lambda$ (in units of~${1 \over h}$) as follows: $1, \Lambda = 0.667$; $2, 
\Lambda = 3.333$; $3, \Lambda = 6.667$; $4, \Lambda = 10.0$; $5, \Lambda = 
13.333$. Each distribution is computed from a set of 2000 data points.}  
\label{fig3}
\end{figure}

	In Fig.~\ref{fig3} we show the normalized distribution of concentration
of $X$ species on Poincar\'e section ${\cal P}$ for different values $\Lambda$. 
The value of $\kappa_2$ is $1.4$; corresponding mass-action law dynamics is 
period-1. 

	One observes that the peak of the distribution shifts toward lower 
concentrations as $\Lambda$ decreases, indicating that the trajectory shrinks 
as the rate diffusion is decreased. That this shrinking is due to a parametric 
shift and not to simple rescaling of volume can be seen from the fact that the 
shift of the peak vanishes as the diffusion becomes faster and the automaton 
trajectory approaches that of the mass-action law. Note that the width of the 
distribution does not vanish as the rate of diffusion increases, but instead 
approaches a non-zero limiting value. This effect arises as a result of fluctuations 
which the global average concentrations experience due to the finite number of 
particles involved in the simulations.

\begin{figure}[htbp]
\begin{center}
\leavevmode
\epsffile{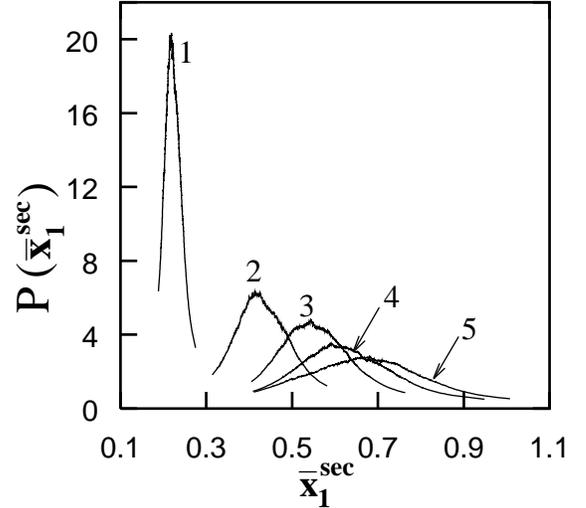}
\end{center}
\caption{Distribution of concentration of $X$ species on Poincare section \protect
${\cal P}$ for different values of $\Lambda$. The mass-action law dynamics is 
period-2 ($\kappa_2 = 1.5$). The graphs are numbered according to the value of 
$\Lambda$ as in Fig.~\protect \ref{fig3}. Each distribution is computed from a 
set of 2000 data points.}  
\label{fig4}
\end{figure}

	Fig.~\ref{fig4} presents similar distributions computed for $\kappa_2 =
1.5$ (the mass-action law dynamics is period-2). The values of $\Lambda$ used 
in these calculations are the same as in Fig.~\ref{fig3}.

Note that for the highest value of $\Lambda$ (i.e. the fastest diffusion) 
the width of distribution in Fig.~\ref{fig4} is considerably greater than in 
Fig.~\ref{fig3}. The additional spreading comes from the fact that the dynamics 
is already beyond the period-doubling bifucation at this value of $\Lambda$. 
However, the concentration fluctuations destroy the bimodality in distribution, 
and we are unable to distinguish the loops of period-2 trajectory even for 
relatively fast diffusion.  Larger system sizes are necessary in order to 
resolve this structure.
	
\begin{figure}[htbp]
\begin{center}
\leavevmode
\epsffile{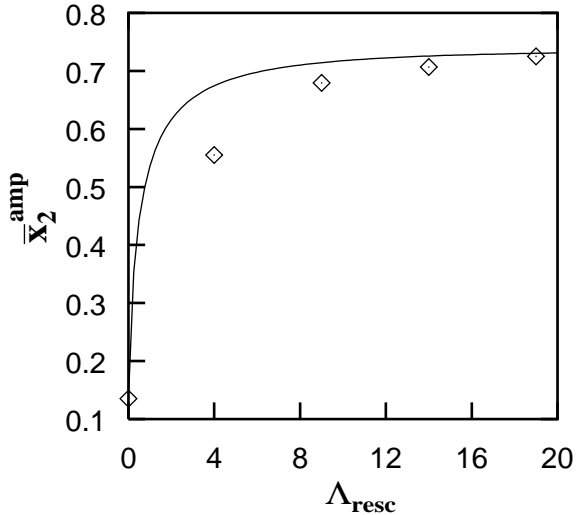}
\end{center}
\caption{Oscillation amplitude for the concentration of $Y$ species in period-1 
regime for different values of $\Lambda$ obtained from the kinetic equations
(\protect \ref{WR_global_ev_a}) -- (\protect \ref{WR_global_ev_c}) (solid line)
and from lattice gas automaton simulations (open diamonds).}
\label{fig5}
\end{figure}

	Finally, in Fig.~\ref{fig5} we show the average oscillation amplitude 
for the concentration of $Y$ species for a series of lattice gas automaton simulations.
The values of the rate constants correspond to period-1 regime in the mass-action
law dynamics and are the same for all simulations, while $\Lambda$ is varied from 
$0.667$ to $13.333$. The results obtained from the kinetic equations 
(\ref{WR_global_ev_a}) -- (\ref{WR_global_ev_c}) are shown by the solid line. 
The oscillation amplitude $\overline{x}_i^{{\rm amp}}$ is defined as the difference between 
the maximum and the minimum of $\overline{x}_i$ over one full cycle of trajectory around 
the unstable focus. Since the present theory is approximate, our primary aim 
here is to present a qualitative, rather than quantitative, comparison of 
theoretical data and simulations. Therefore, the exponent $\Lambda$ is given 
in Fig.~\ref{fig5} in rescaled form, $\Lambda_{{\rm resc}} = {\Lambda - 
\Lambda_{{\rm ref}} \over \Lambda_{{\rm ref}}}$. Here $\Lambda_{{\rm ref}}$ is 
the reference value of $\Lambda$, which is equal to $0.667$ for the automaton 
simulations, and $0.128$ for the theoretical data; at these values of $\Lambda$ 
the amplitude obtained from simulations and predicted theoretically match 
exactly. 

One sees that, qualitatively, kinetic equations (\ref{WR_global_ev_a}) -- (\ref{WR_global_ev_c})
correctly describe the breakdown of the mass-action law description for the 
dynamics of global concentrations as the diffusion coefficient is decreased. 

\section{Conclusions}
\label{conclusions}
	In this paper, the breakdown of the mean-field dynamics in 
non-equilibrium reaction-diffusion systems is studied by the means of a 
Chapman-Enskog type expansion using a probabilistic discrete-space, discrete-time
model. It is shown that this breakdown is related to the memory effects in the
evolution of average concentrations that emerge for finite values of diffusion 
coefficient. The memory kernel, although quite complicated in general case, 
was simplified for the case of Willamowski-R\"ossler reaction-diffusion system 
to a simple exponential $e^{-\Lambda (t - t')}$, where the exponent $\Lambda$ 
is determined by the diffusion coefficient and $t$ is the (continuous) time 
variable.

	A numerical analysis of theoretically derived kinetic equations with
memory led to a further insight into the nature of the breakdown of mean-field 
dynamics. Namely, the effect of a decreasing diffusion coefficient on the reaction 
dynamics can be described by a backward shift in the bifurcation cascade of 
the mass-action law, i.e. the breakdown is essentially a parametric phenomenon. 
This conclusion is further supported by a comparison to the results of lattice 
gas automaton simulations of Willamowski-R\"ossler model. 				 

	The breakdown of mean-field description for the dynamics of reaction-diffusion 
systems has been considered in a number of earlier studies. Thus, simulations 
of the reaction-diffusion master equation for Brusselator model showed that 
the deterministic limit cycle is gradually destroyed as the rate of diffusion 
is decreased and the inhomogeneous dynamic modes become excited \cite{baras}. 
Recently, Bussemaker and Brito used ring kinetic theory to explain shrinking 
of the limit cycle for Maginu model with attenuation in the rate of 
diffusion~\cite{bussemaker}. 
These authors were able to qualitatively reproduce the effect of breakdown of 
the mean-field kinetics observed in lattice gas automaton simulations by 
accounting for the reactive correlations in a phenomenological way. Because the
probabilistic model used in these studies is similar to that we used here, it 
may be interesting to consider the relation among these approaches.

	There is a number of important questions which were not considered 
in this work. For instance, development of an analytic means for estimating 
the effective parametric shift exhibited by the solutions of kinetic equations 
(\ref{WR_global_ev_a}) -- (\ref{WR_global_ev_c}) as a function of diffusion 
coefficient would be useful. A different, no-less-challenging question 
concerns the extension of this theory to the level of local concentrations and the 
incorporation of long wavelength diffusion
modes into the description of the dynamics. Such an extension would clearly be
expedient for studies of stability and evolution of spatio-temporal structures 
in reaction-diffusion systems. 

\section*{Acknowledgments}	 

	This work was partially supported by Natural Sciences and Engineering 
Research Council of Canada. M.V. also benefits from a Connaught Scholarship.
Computing resources for this study were provided in part by the University of 
Toronto Information Commons.

\section*{Appendix A}

	Substituting $P_1({\bf x}({\bf r}),n)$ from (\ref{p1}) into the r.h.s.
of the first of the solvability conditions (\ref{solv_cond}) gives:
\wdtxt
\setcounter{equation}{0}
{\renewcommand{\theequation}{A.\arabic{equation}}
\begin{equation}				
\label{p1_1_solv}
\Bigl \langle P_1({\bf x}({\bf r}),n) \Bigr \rangle = \sum_{n'=0}^{n-1}
\left \langle \left( \hat{I} + \hat{W}^D \right)^{n-n'-1} \left[
\hat{W}^R - {\bf K}(\overline{{\bf x}}({\bf r},n')) {\partial \over 
\partial \overline{{\bf x}}} \right] P_B({\bf x}({\bf r}),n') \right 
\rangle.
\end{equation}		   
\nrtxt
To simplify this expression, note that by definition: 
\begin{equation}
\label{WD_power}
\left( \hat{I} + \hat{W}^D \right)^p \equiv \hat{I} + \sum_{p'=1}
^p \sum_{\mbox{\boldmath $\pi$\unboldmath $(p')$}}^{} \left( {\hat{W}^D} 
\right)^{p'},
\end{equation}
where $p$ is any positive integer and $\sum_{\mbox{\boldmath $\pi$\unboldmath 
$(p')$}}^{} \left( {\hat{W}^D} \right)^{p'}$ denotes sum over all possible 
time-ordered products of $p'$ operators $\hat{W}^D$ selected from the 
full set of $p$ such operators.

	Given (\ref{WD_power}), we can write (\ref{p1_1_solv}) as:
\wdtxt
\begin{eqnarray}
\label{p1_1_solv_cont}
\Bigl \langle P_1({\bf x}({\bf r}),n) \Bigr \rangle &=& \sum_{n'=0}^{n-1}
\Biggl \{ \left \langle \left[ \hat{W}^R - {\bf K}(\overline{{\bf x}}
({\bf r},n')) {\partial \over \partial \overline{{\bf x}}} \right] P_B(
{\bf x}({\bf r}),n') \right \rangle \Biggr. \\
& & \Biggl. \hskip 0.75 cm 
+ \sum_{l=1}^{n-n'-1} \sum_{\mbox{\boldmath $\pi$\unboldmath $(l)$}}^{} \left 
\langle \left( {\hat{W}^D} \right)^l \left[ \hat{W}^R - {\bf K}
(\overline{{\bf x}}({\bf r},n')) {\partial \over \partial \overline{{\bf x}}} 
\right] P_B({\bf x}({\bf r}),n') \right \rangle \Biggr \}. 
\nonumber
\end{eqnarray}			       
\nrtxt
The first summand in (\ref{p1_1_solv_cont}) vanishes since for all $n$ 
we have:
\begin{eqnarray}
\label{auxiliary_1}
\Bigl \langle \hat{W}^R P_B({\bf x}({\bf r}),n) \Bigr \rangle &=& 0,\nonumber \\
\Bigl \langle P_B({\bf x}({\bf r}),n) \Bigr \rangle &=& 1.
\end{eqnarray}
The remainder of our analysis is easier to carry out in matrix form. We define:
\begin{equation}
\label{B-matrix}
B_{{\bf x}({\bf r})}^{(l)}(n) = \left( \hat{W}^D \right)^l \hat{S} 
P_B({\bf x}({\bf r}),n), 
\end{equation} 
where
\begin{eqnarray}
& &\hat{S} = \hat{W}^R - {\bf K}(\overline{{\bf x}}({\bf r},n)) 
{\partial \over \partial \overline{{\bf x}}}, \\
& &\left( \hat{W}^D \right)^0 \equiv \hat{I}, \nonumber 
\end{eqnarray}
and re-write (\ref{p1_1_solv_cont}) in terms of vector elements 
$B_{{\bf x}({\bf r})}^{(l)}(n)$ as follows:
\begin{equation}
\label{p1_1_solv_matr}		       
\hspace{-14 pt}
\Bigl \langle P_1({\bf x}({\bf r}),n) \Bigr \rangle = \sum_{n'=0}^{n-1} 
\sum_{l=1}^{n-n'-1} \sum_{\mbox{\boldmath $\pi$\unboldmath $(l)$}}^{} 
\left \langle {\hat{W}^D} B_{{\bf x}({\bf r})}^{(l-1)}(n') \right
\rangle.
\end{equation}

	Using the definition of the matrix elements of the diffusion evolution 
operator given by (\ref{Diff_chain}) and (\ref{Wdiff}), we expand the leftmost 
$\hat{W}^D$ in (\ref{p1_1_solv_matr}) and obtain:
\wdtxt
\begin{eqnarray}
\label{p1_1_solv_final}
\Bigl \langle P_1(x({\bf r}),n) \Bigr \rangle &=& \sum_{n'=0}^{n-1} 
\; \sum_{l=1}^{n-n'-1} \sum_{\mbox{\boldmath $\pi$\unboldmath $(l)$}}^{} 
\left \langle \hat{W}^D B_{x({\bf r})}^{(l-1)}(n') \right \rangle, \nonumber 
\\ 
& & \\
&=& 0. \nonumber
\end{eqnarray}
\nrtxt
Proceeding in the similar manner with the other condition in (\ref{solv_cond}),
we find:
\wdtxt
\begin{eqnarray}
\label{p1_2_solv}
\Bigl \langle {\bf x}({\bf r}) P_1({\bf x}({\bf r}),n) \Bigr \rangle &=& 
\sum_{n'=0}^{n-1} \left \langle {\bf x}({\bf r}) \left( \hat{I} + \hat{
{\bf W}}^D \right)^{n-n'-1} \hat{S} P_B({\bf x}({\bf r}),n') \right 
\rangle, \\
&=& \sum_{n'=0}^{n-1} \left \{ \left \langle {\bf x}({\bf r}) \hat{S}
P_B({\bf x}({\bf r}),n') \right \rangle 
+ \sum_{l=1}^{n-n'-1} \sum_{\mbox{\boldmath $\pi$\unboldmath 
$(l)$}}^{} \left \langle {\bf x}({\bf r}) \left( {\hat{W}^D} \right)^l 
\hat{S} P_B({\bf x}({\bf r}),n') \right \rangle \right \}. \nonumber
\end{eqnarray}			       
\nrtxt
\parbox[b]{8.5 cm}{
The summand on the second line of (\ref{p1_2_solv}) again vanishes 
because of the following relations:
\begin{eqnarray}
\label{auxiliary_2}
\Bigl \langle x_i({\bf r}) \hat{W}^R P_B({\bf x}({\bf r}),n) 
\Bigr \rangle &=& K_i(\overline{{\bf x}}({\bf r},n)), \nonumber \\
\Bigl \langle x_i({\bf r}) P_B({\bf x}({\bf r}),n) \Bigr \rangle &=& 
\overline{x}_i({\bf r},n).
\end{eqnarray}}
Hence, one can again expand the leftmost operator $\hat{W}^D$ and re-write 
(\ref{p1_2_solv}) as follows:
\vspace{36 pt}
\wdtxt	       
\begin{eqnarray}
\label{p1_2_solv_cont}
\Bigl \langle {\bf x}({\bf r}) P_1({\bf x}({\bf r}),n) \Bigr \rangle &=& 
\sum_{n'=0}^{n-1} \; \sum_{l=1}^{n-n'-1} \sum_{\mbox{\boldmath $\pi$\unboldmath 
$(l)$}} \left \langle {\bf x}({\bf r}) \hat{W}^D B^{(l-1)}_{{\bf x}(
{\bf r})}(n') \right \rangle, \nonumber \\
\\
&=& {1 \over N} \sum_{n'=0}^{n-1} \; \sum_{l=1}^{n-n'-1} \sum_{\mbox{\boldmath 
$\pi$\unboldmath $(l)$}} \left \langle \left[ \chi({\bf r},n'') - x({\bf r}) 
\right] B_{x({\bf r})}^{(l-1)}(n') \right \rangle, \nonumber 
\end{eqnarray}			       
\nrtxt
where $n''$ is the moment of time at which the leftmost operator $\hat{W}^D$
was applied ($0 \leq n'' \leq n-n'-1$$), B_{x({\bf r})}^{(l)}(t)$ is defined 
by (\ref{B-matrix}). 

	One can now separate the leftmost operator $\hat{W}^D$ from 
vector elements $B_{x({\bf r})}^{(l)}(n)$ where possible, and repeat the 
calculations in (\ref{p1_2_solv_cont}). Taking advantage of the results in
(\ref{auxiliary_1}) and (\ref{auxiliary_2}), we obtain:
\wdtxt
\begin{eqnarray}
\label{p1_2_solv_final}
\Bigl \langle {\bf x}({\bf r}) P_1({\bf x}({\bf r}),n) \Bigr \rangle &=& 
{1 \over N} \sum_{n'=0}^{n-1} \; \sum_{l=1}^{n-n'-1} \sum_{\mbox{\boldmath 
$\pi$\unboldmath $(l)$}} \left \langle \left[ \chi({\bf r},n'') - x({\bf r}) 
\right] B_{x({\bf r})}^{(l-1)}(n') \right \rangle, \nonumber \\
&=& -{1 \over N} \sum_{n'=0}^{n-1} \left \{ (n-n'-1) \left \langle x({\bf r}) 
B_{x({\bf r})}^{(0)}(n') \right \rangle + \sum_{l=2}^{n-n'-1} \sum_
{\mbox{\boldmath $\pi$\unboldmath $(l)$}} \left \langle x({\bf r}) 
B^{(l-1)}_{{\bf x}({\bf r})}(n') \right \rangle \right \}, \nonumber \\
&=& -{1 \over N} \sum_{n'=0}^{n-1} \left \{ (n-n'-1) \left \langle x({\bf r}) 
B_{x({\bf r})}^{(0)}(n') \right \rangle + \sum_{l=2}^{n-n'-1} \sum_
{\mbox{\boldmath $\pi$\unboldmath $(l)$}} \left \langle x({\bf r}) 
\hat{W}^D B^{(l-2)}_{{\bf x}({\bf r})}(n') \right \rangle 
\right \}, \\
& & \nonumber \\
& & \hskip 4 cm \dots \;\; , \nonumber \\				      
& & \nonumber \\
&=& \sum_{n'=0}^{n-1} \; \sum_{l=1}^{n-n'-1} {n-n'-1 \choose l} \left( - {1 
\over N} \right)^l \left \langle x({\bf r}) B_{x({\bf r})}^{(0)}(n') 
\right \rangle, \nonumber \\
& & \nonumber \\
&=& 0. \nonumber
\end{eqnarray}		
}
\nrtxt

\section*{Appendix B}

	It is convenient to consider the continuous-time limit of kinetic 
eqns (\ref{WR_global_ev_a}) -- (\ref{WR_global_ev_c}) in rescaled time 
units, $h' = \gamma h$. Thus, we define a discrete, real-valued time variable 
$t_n = {n \over \gamma} h'$, and re-cast (\ref{WR_global_ev_a}) -- (\ref{WR_global_ev_c}) 
in the following form:					     
\vfill 
\pagebreak
\wdtxt
\setcounter{equation}{0}
{\renewcommand{\theequation}{B.\arabic{equation}}
\begin{eqnarray}
\label{WR_cont_time_a}				   
\overline{x}_1(t_n+{h' \over \gamma}) - \overline{x}_1(t_n) &=& h' R_1(\overline{{\bf x}}(t_n)) 
\nonumber \\
&-& {h'}^2 {N \over {N-1}} \sum_{n'=0}^{n-1} \left[ \left( 1 - \lambda \right)^
{{t_n-t_{n'}-{h'/\gamma} \over {h'/\gamma}}} {\kappa_{-1} C_{x_1,x_1}(t_{n'}) 
- \kappa_{-2} C_{x_2,x_2}(t_{n'}) \over {h'/\gamma}} \right] \\ 
&-& {h'}^2 \sum_{n'=0}^{n-1} \left[ \left( 1 - \lambda \right)^{{t_n-t_{n'}-
{h'/\gamma} \over {h'/\gamma}}} {\kappa_2 C_{x_1,x_2}(t_{n'}) + \kappa_4 
C_{x_1,x_3}(t_{n'}) \over {h'/\gamma}} \right], \nonumber \\
\nonumber \\
\label{WR_cont_time_b}				   
\overline{x}_2(t_n+{h' \over \gamma}) - \overline{x}_2(t_n) &=& h' R_2(\overline{{\bf x}}(t_n)) 
\\&-& {h'}^2 \sum_{n'=0}^{n-1} \left( 1 - \lambda \right)^{{t_n-t_{n'}-{h'/\gamma} 
\over {h'/\gamma}}} \left[{N \over {N-1}} {\kappa_{-2} C_{x_2,x_2}(t_{n'}) 
\over {h'/\gamma}} - {\kappa_{2} C_{x_1,x_2}(t_{n'}) \over {h'/\gamma}} \right], 
\nonumber \\ 
\nonumber \\
\label{WR_cont_time_c}				   
\overline{x}_3(t_n+{h' \over \gamma}) - \overline{x}_3(t_n) &=& h' R_3(\overline{{\bf x}}(t_n)) 
\\&-& {h'}^2  \sum_{n'=0}^{n-1} \left( 1 - \lambda \right)^{{t_n-t_{n'}-{h'/\gamma} 
\over {h'/\gamma}}} \left[ {N \over {N-1}} {\kappa_{-5} C_{x_3,x_3}(t_{n'}) 
\over {h'/\gamma}} + {\kappa_4 C_{x_1,x_3}(t_{n'}) \over {h'/\gamma}} \right],
\nonumber 
\end{eqnarray}
\nrtxt

	We now have to evaluate the limit of (\ref{WR_cont_time_a}) -- (\ref{WR_cont_time_c}) 
as $h'$ tends to $0$ in such a way that the diffusion coefficient remains 
unchanged. Formally, we can do so by letting $\lambda = \Lambda h' = \gamma 
\Lambda h$, and keeping $\Lambda$ constant as $h' \rightarrow 0$. From the 
point of view of statistical theory, such an operation is equivalent to 
replacing the diffusion transition probabilities (\ref{no_corr}) with appropriate 
transition rates (such that the resulting diffusion coefficient is identical 
in both cases) as the diffusion Markov chain (\ref{Diff_chain}) is replaced 
with a continuous-time Markov process. Proceeding in this manner, we obtain:
\wdtxt
\begin{eqnarray}
\label{cont_time_final_a}
{d \overline{x}_1(t) \over d t} = R_1(\overline{{\bf x}}(t)) &-& {N \over N-1} 
\int_0^t e^{- \Lambda (t - t')} \Bigl[ \kappa_{-1} {\cal C}_{x_1,x_1}(t') - 
\kappa_{-2} {\cal C}_{x_2,x_2}(t') \Bigr] d t' \\ 
&-& \int_0^t e^{- \Lambda (t - t')} \Bigl[ \kappa_2 {\cal C}_{x_1,x_2}(t') + 
\kappa_4 {\cal C}_{x_1,x_3}(t') \Bigr] d t', \nonumber \\
\nonumber \\
\label{cont_time_final_b}				   
{d \overline{x}_2(t) \over d t} = R_2(\overline{{\bf x}}(t)) &-& \int_0^t 
e^{- \Lambda (t - t')} \left[ {N \kappa_{-2} \over {N-1}} {\cal C}_{x_2,x_2}(t') 
- \kappa_{2} {\cal C}_{x_1,x_2}(t') \right] d t', \\ 
\nonumber \\
\label{cont_time_final_c}				   
{d \overline{x}_3(t) \over d t} = R_3(\overline{{\bf x}}(t)) &-& \int_0^t 
e^{- \Lambda (t - t')} \left[ {N \kappa_{-5} \over {N-1}} {\cal C}_{x_3,x_3}(t') 
+ \kappa_4 {\cal C}_{x_1,x_3}(t') \right] d t',
\end{eqnarray}					  
}
\nrtxt
where ${\cal C}_{x_i,x_j}(t) = {C_{x_i,x_j}(t) \over {h' /\gamma}} = 
{C_{x_i,x_j}(t) \over h}$, and $C_{x_i,x_j}(t)$ are defined by (\ref{c_def}) and 
(\ref{WR_mem_func}). Eqns (\ref{cont_time_final_a}) -- (\ref{cont_time_final_c}) 
represent the continuous-time form of the kinetic equations (\ref{WR_global_ev_a}) -- (\ref{WR_global_ev_c}). 
The parameter $\Lambda$ is defined as $\Lambda = {\lambda \over h'} = {2 \over 
\nu N \gamma h}$, where $h$ is the time unit of diffusion Markov chain 
(\ref{Diff_chain}).

\end{multicols}

\begin{thebibliography}{99}			      

\vspace{-48 pt}
\bibitem{gen}    A.S. Mikhailov,  {\em   Foundations   of
Synergetics      I.      Distributed      Active     Systems},
(Springer-Verlag,  Berlin,  1994).

\bibitem{sokolov}  I.M. Sokolov and A. Blumen, Europhys. Lett., {\bf 27},
495 (1994).

\bibitem{felderhof}  B.U. Felderhof and R.B. Jones, J. Chem. Phys., {\bf 103},
10201 (1995); {\it ibid.}, {\bf 106}, 954 (1997).

\bibitem{privman} {\em Nonequilibrium Statistical Mechanics in One Dimension},
ed. V. Privman, (Cambridge University Press, Cambridge, 1997).

\bibitem{broeck} C. Van den Broeck, J. Houard, M. Malek Mansour, Physica,
{\bf 101A}, 167 (1980).

\bibitem{wu}  X.-G.  Wu  and  R.  Kapral,  J. Chem. Phys. 
{\bf 100}, 5936 (1994); X.-G.  Wu  and  R.  Kapral, Phys. Rev. Lett., {\bf 70},
1940 (1993).

\bibitem{baras}  F. Baras, Phys. Rev. Lett. {\bf 77}, 1398 (1996).

\bibitem{nicolis}  P. Geysermans and G. Nicolis, J. Chem. Phys., {\bf 99},
8964 (1993); P. Peeters and G. Nicolis, Physica A, {\bf 188}, 426 (1992).

\bibitem{uhlenbeck}  G. Uhlenbeck, {\em Lectures in Statistical Mechanics},
(Am. Math. Soc., Providence, RI, 1963); S. Chapman and T. G. Cowling,
{\em Mathematical Theory of Non-Uniform Gases}, (Cambridge U. P. ,
London, 1961).

\bibitem{wr}  K.-D.  Willamowski  and  O.E.  R\"{o}ssler,  Z. Naturforsch. 
{\bf 35a}, 317 (1980).

\bibitem{cox}  D.R. Cox and H.D. Miller, {\em The Theory of Stochastic 
Processes}, (Methuen, London, 1965). 

\bibitem{malevan}  A. Malevanets and R. Kapral, Phys. Rev. E., {\bf 55},
5657 (1997).

\bibitem{LGA} J.-P. Boon, D. Dab, R. Kapral and A. Lawniczak, Phys. Rep., 
{\bf 273}, 55 (1996).

\bibitem{modce} R. Kapral, S. Hudson and J. Ross, J. Chem. Phys. {\bf
53}, 4387 (1970).
			      
\bibitem{scott}  S.K. Scott, {\em Chemical Chaos}, (Oxford University Press,
New York, 1991).

\bibitem{aguda}  B.D. Aguda and B.L. Clarke, J. Chem. Phys., {\bf 89}, 7428 (1988).

\bibitem{bussemaker}  H.J. Bussemaker and R. Brito, J. Stat. Phys., {\bf 87},
1165 (1997).

\end{thebibliography}
\end{document}